\documentclass[11pt]{article}
\usepackage[intlimits]{amsmath}\usepackage{amssymb}\usepackage{xspace}
\usepackage[dvips]{epsfig}
\usepackage{slashbox}
\def\lb{\label}
\def\be{\begin{equation}}
\def\ba{\begin{eqnarray}}
\def\ea{\end{eqnarray}}
\def\ds{\displaystyle}
\def\ol{\overline}
\def\bb{\bibitem}
\def\p{\partial_}
\def\e{{\rm e}}
\def\bt{\begin{tabular}}
\def\et{\end{tabular}}

\begin{document}

\title{
   \begin{flushright} \begin{small}
     LAPTH-929/02  \\ DTP-MSU/02---21 \\
  \end{small} \end{flushright}
{\bf Linear dilaton black holes} }
\author{
{\bf G\'erard Cl\'ement$^{a}$\thanks{ Email: gclement@lapp.in2p3.fr},
Dmitri Gal'tsov$^{a,b}$
\thanks{Supported by RFBR; Email:  galtsov@grg.phys.msu.ru}
and C\'edric Leygnac$^{a}$\thanks{ Email: leygnac@lapp.in2p3.fr}} \\ \\
$^{a}$Laboratoire de  Physique Th\'eorique LAPTH (CNRS), \\
B.P.110, F-74941 Annecy-le-Vieux cedex, France\\ $^{b}$Department
of Theoretical Physics, Moscow State University,\\ 119899, Moscow,
Russia,}

\date{30 August 2002}
\maketitle
\begin{abstract}
We present new solutions to Einstein-Maxwell-dilaton-axion (EMDA)
gravity in four dimensions describing black holes which asymptote to
the linear dilaton background. In the non-rotating case they can be
obtained as the limiting geometry of dilaton black
holes. The rotating solutions (possibly endowed with a NUT
parameter) are constructed using a generating technique based on the
$Sp(4,R)$ duality of the EMDA system. In a certain limit (with no event
horizon present) our rotating solutions coincide with supersymmetric
Israel-Wilson-Perj\`es type dilaton-axion solutions. In the presence of an
event horizon supersymmetry is broken. The temperature of the static
black holes is constant, and their mass does not depend on it, so the
heat capacity is zero. We investigate geodesics and wave propagation in
these spacetimes and find superradiance in the rotating case. Because 
of the non-asymptotically flat nature of the geometry, certain modes 
are reflected from infinity; in particular, all superradiant modes are 
confined. This leads to a classical instability of the rotating solutions.
The non-rotating linear dilaton black holes are shown to be stable 
against spherical perturbations.
\end{abstract}

\newpage
\section{Introduction}
Non-asymptotically flat spacetimes containing event horizons recently
attracted attention in connection with the conjecture of AdS/CFT
correspondence and its generalizations. In most cases the asymptotic
geometry considered was of the AdS type, such
configurations frequently demonstrate holography, the field theory
side being some conformal field theory. As is well-known, AdS
geometry typically arises in the near-horizon limit of non-dilatonic
BPS black holes and/or p-branes. Other examples of holographic
backgrounds are provided by the linear dilaton solutions which arise
as near-horizon limits of  {\em dilatonic} holes/branes 
\cite{AhBeKuSe98}. In such
cases the dual theory is likely to be not a local field theory. 
In particular, the near-horizon limit of parallel
NS5 branes is dual to the decoupled NS5 brane theory known as
``little string theory'' \cite{Se97,Ah99}. It is therefore interesting to
investigate other backgrounds which asymptote to the linear dilaton
vacuum.

Here we present new four-dimensional solutions describing rotating
black holes in a linear dilaton background. Actually, a non-rotating
version of such a black hole was given long ago by Giddings and
Strominger \cite{GS} (see also \cite{KP}) as some limit of extremal
dilaton black holes \cite{gibbons82,GM,GHS,KP} (`horizon plus throat' 
geometry), but apparently
was not fully appreciated as an exact solution of 
Einstein-Maxwell-dilaton theory describing a non-asymptotically flat
black hole. We find that within the Einstein-Maxwell-dilaton-axion
theory there exist rotating black hole solutions (possibly with NUT
parameter) which asymptote to the linear dilaton space-time. The
latter is supersymmetric ($1/2$ BPS), while our black holes break all
supersymmetries. Presumably their higher dimensional counterparts
exist as well and may serve as an arena  for non-supersymmetric
holography, although we do not present any evidence for this here,
being concentrated on geometric and other physical properties of  
the new solutions in four dimensions.

In the non-rotating case our black holes can be regarded as
excitations over the linear dilaton vacua, forming a two-parameter
family, which can be obtained as near-horizon and near-extreme limits 
of the ``parent''
dilaton black holes. To identify the physical mass we use the
Brown-York formalism \cite{BY}. These spacetimes are not asymptotically
flat, so a suitable substraction procedure is needed (of the type
described in \cite{HaHo95}) to obtain finite quantities from the
integrals divergent at infinity. For the non-rotating black hole the
linear dilaton background is a natural (and unique) choice for such
substraction background. With this prescription we are able to establish
the validity of the first law of black hole thermodynamics $d{\cal
M}=TdS$, in which the Hawking temperature is independent of the black
hole mass, but depends on the parameter of the background. The
entropy is given by the quarter of the area as usual. The mass is
therefore independent on the temperature, so the heat capacity is
precisely zero. In other words, these black holes will be in a marginal
thermodynamical equilibrium with the heat bath. We also 
check that the non-rotating linear dilaton
black holes are classically stable against spherical perturbations.

We also present a rotating version (possibly with NUT) of linear
dilaton black holes, in which case the axion is non-zero, but
(as well as the dilaton) is independent of the black hole mass. These
may be regarded as excitations over the rotating 
Israel-Perj\`es-Wilson (IWP) \cite{Ka94} solutions  which correspond to a
vanishing black hole mass parameter, but non-zero rotation parameter.
The metric possesses an ergosphere outside the event horizon, which
rotates with some angular velocity $\Omega_h$. The angular momentum
$J$ computed as a surface integral at spatial infinity contains both
a geometrical contribution and that coming from the Maxwell field.
The Hawking temperature and the entropy depend both on the mass and
the angular momentum of the black hole, these quantities are shown to satisfy
the first law in the form $d{\cal M}=TdS+\Omega_hdJ$.  

Surprisingly, these new rotating black holes are related in a 
certain way to the five-dimensional Myers-Perry \cite{MP5} 
rotating black hole with two equal angular
momenta. This relation is non-local and derives from the recently
found new classical duality between the $D=4$ EMDA system and 
six-dimensional vacuum gravity \cite{bremda} (for more general  
non-local dualities of this kind see \cite{CGS}). The symmetries of 
the six-dimensional solution combine in a non-trivial way the space-time
symmetry of the four-dimensional EMDA solution with the internal symmetry
$SO(2,3)\sim Sp(4,R)$ of the stationary EMDA system.

The Hamilton-Jacobi equation in the rotating linear dilaton black
hole spacetime is shown to be separable as in the Kerr case, 
indicating the
existence of a Stachel-Killing tensor. Timelike geodesics do not
escape to infinity, while null ones do so for an infinite affine
parameter. The Klein-Gordon equation is also separable, the mode
behavior near the horizon exhibits the superradiance phenomenon. 
We show that all
superradiant modes are reflected at large distances from the black
hole, so there is no superradiant flux at infinity, in contrast with
the case of massless fields in the Kerr spacetime. This situation 
is similar to the confinement of {\em massive}
superradiant modes in the Kerr metric, which are reflected back to the
horizon causing stimulated emission and absorption. In the
classical limit this leads to an amplification effect due to
the positive balance between emission and absorption. 
In the Kerr spacetime this phenomenon is rather small, being present 
only for massive modes. In our case massless modes are confined too, 
therefore all superradiant modes will form a cloud outside 
the horizon with an exponentially growing amplitude. 
This is in fact a classical instability
which manifests itself in the rapid transfer of angular momentum
from the hole to the outside matter cloud. A similar conclusion 
was obtained in \cite{HaRe99} for the case of the Kerr-AdS spacetime 
with reflecting boundary conditions on the AdS boundary.

For certain values of the parameters our solutions represent a
naked singularity on the rotating linear dilaton background. We have
found that in some cases the spectrum of the massless scalar field 
may be partially or entirely discrete. 

The plan of the paper is as follows. In Sect.~2 we investigate the
near-horizon limit of static dilaton black holes and show that
a suitable limiting procedure leads to a two-parameter family of
exact solutions of the EMDA theory, which includes static black holes
asymptoting to the linear dilaton background. In Sect.~3 we apply
the $Sp(4,R)$ generating technique to find rotating and NUT
generalizations of the above solutions, forming now a five-parameter
family. 
Then in Sect.~4 we apply a non-local uplifting precedure to
find six and five-dimensional counterparts to our rotating
solution without NUT parameter, and show that it corresponds to the
five-dimensional vacuum Kerr solution with two equal rotation
parameters. The next Sect.~5 is devoted to the determination of the 
physical mass, angular momentum and other parameters of our new black
holes, and to the demonstration of the validity of the first black hole law.
In Sect.~6 we investigate geodesics and modes of a minimally
coupled scalar field showing, in particular, confinement of the
superradiant modes. The stability of the static solutions against
spherically symmetric perturbations is established in Sect.~7.

\section{Static linear dilaton black holes}
\setcounter{equation}{0}

Consider the EMDA theory arising as a
truncated version of the bosonic sector of $D=4,\;{\cal N}=4$
supergravity. It comprises the dilaton $\phi$ and (pseudoscalar) axion 
$\kappa$ coupled to an Abelian vector field $A_{\mu}$ : \begin{equation} \label{ac} S =
\frac{1}{16\pi}\int d^4x\sqrt{|g|}\left\{-R +
2\partial_\mu\phi\partial^\mu\phi + \frac{1}{2}\e^{4\phi}
{\partial_\mu}\kappa\partial^\mu\kappa
-\e^{-2\phi}F_{\mu\nu}F^{\mu\nu}-\kappa F_{\mu\nu}{\tilde
F}^{\mu\nu}\right\} , \end{equation} where\footnote{Here
$E^{\mu\nu\lambda\tau} \equiv
|g|^{-1/2}\varepsilon^{\mu\nu\lambda\tau}$, with $\varepsilon^{1234}
= +1$, where $x^4 = t$ is the time coordinate.} ${\tilde
F}^{\mu\nu}=\frac{1}{2}E^{\mu\nu\lambda\tau}F_{\lambda\tau},\;
F=dA\;$. The black hole solutions to this theory were extensively
studied in the recent past
\cite{Ka92,Ka94,GaKe94,GaKe96,BeKaOr96,ClGa96}.

In this section we will study a solution, describing charged
static dilaton black holes, found in Refs.
\cite{gibbons82,GM,GHS}. Recall the electrically charged
solution (whose magnetic dual is obtained by the
replacement $(\phi, F) \to (\hat{\phi} = -\phi, \hat{F} = \e^{-2\phi}
\tilde{F})$): 
\ba
ds^2 & = & (1-\frac{r_+}{r}) dt^2-(1-\frac{r_+}{r})^{-1}
dr^2-r^2(1-\frac{r_-}{r}) d\Omega^2,\label{gGHS}\\
\e^{2 \phi}& = &
\e^{2\phi_\infty}(1-\frac{r_-}{r}),\qquad\kappa=0,
\label{phiGHS}\\
F & = & \frac{\ol{Q}{\ds\e^{\phi_\infty}}}{r^2} dr\wedge dt. \label{FGHS}
\ea where $\phi_\infty$ is the asymptotic value of the dilaton field. 
The mass and charge of the black hole are
\begin{equation}
M=\frac{r_+}{2}, \quad Q=e^{-\phi_{\infty}}\ol{Q}=e^{-\phi_{\infty}}
\sqrt{\frac{r_+ r_-}{2}}.
\end{equation}

The extreme black hole ($r_+=r_-$) is singular, the singularity being
marginally trapped. In \cite{KP} and \cite{GS} (the ``infinite
throat with linear dilaton'' case) the near-horizon limit of the
extreme dilaton black hole has been considered, leading to \ba
ds^2 & = & \frac{\rho}{\sqrt{2}\,\ol{Q}}
dt^2-\frac{\sqrt{2}\,\ol{Q}}{\rho}(d\rho^2
+\rho^2 d\Omega^2),\label{gb0}\\
\e^{2 \phi}& =&\e^{2\phi_\infty}\frac{\rho}{\sqrt{2}\,\ol{Q}}, \qquad
F=\frac{1}{2\,\ol{Q}}\e^{\phi_\infty} d\rho\wedge dt. \label{phiFb0} \ea
The string metric associated with the magnetic version of
(\ref{gb0})-(\ref{phiFb0}) is 
\begin{equation}\label{str0} ds_{str}^2 =
\e^{2\hat{\phi}}ds^2 = \e^{-2\phi_{\infty}}\bigg[dt^2 - dw^2 -
2\ol{Q}^2\,d\Omega^2\bigg], \end{equation} with $w = \sqrt{2}\,\ol{Q}\ln\rho$. This
spacetime is a cylinder $R^2\times S^2$, and the associated
dilaton \be \hat{\phi} = -\frac{w}{\sqrt{2}\,\ol{Q}} + const. \end{equation} is
linear.

Returning to the Einstein metric (\ref{gb0}), we see from the
conformal map (\ref{str0}) that $\rho=0$ ($w = -\infty$) is a
null bifurcate singularity, while $\rho=\infty$ ($w = + \infty$)
is at spacelike and null infinity. While all curvature invariants
go to zero at spacelike infinity, this metric is not
asymptotically flat. In fact the following coordinate
transformation
\begin{equation}
\xi=\alpha^{-1}\sqrt{\rho}\cosh(2\alpha^2 t)\,,\quad
\eta=\alpha^{-1}\sqrt{\rho}\sinh(2\alpha^2 t) \end{equation} (with
$\alpha^{-2}=4\sqrt{2}\,\ol{Q}$) transforms (\ref{gb0}) to
\begin{equation}
ds^2=d\eta^2-d\xi^2-\frac{\xi^2-\eta^2}{4}d\Omega^2\,.
\end{equation}
When $\xi$ goes to infinity with $\eta$ held fixed this last
metric asymptotes to the conical spacetime
\begin{equation}
ds^2=d\eta^2-d\xi^2-\frac{\xi^2}{4}d\Omega^2\,.
\end{equation}

Now we shall generalize the ``linear dilaton vacuum'' metric
(\ref{gb0}) by studying the near-horizon and near-extremal limit
of (\ref{gGHS})-(\ref{FGHS}). Putting
\begin{equation}
\quad r_-=\epsilon^{-1}r_0,  \quad r_+=\epsilon^{-1}r_0+\epsilon
b, \quad r= \epsilon^{-1}r_0+\epsilon\rho, \quad t =
\epsilon^{-1}\bar{t},
\end{equation}
where the dimensionless parameter $\epsilon$ shall eventually be
taken to zero, the solution (\ref{gGHS})-(\ref{FGHS}) becomes \ba
ds^2&=&\frac{\rho-b}{r_0+\epsilon^2\rho}d\bar{t}^2-
\frac{r_0+\epsilon^2\rho}{\rho-b}
d\rho^2-\rho(r_0+\epsilon^2\rho)d\Omega^2,\\
\e^{2\phi}&=&\frac{\rho}{r_0+\epsilon^2\rho}, \qquad F =
\frac{\epsilon \ol{Q}}{(r_0+\epsilon^2\rho)^2}\,d\rho\wedge d\bar{t},
\ea where we have chosen $\phi_\infty=-\ln \epsilon$. Finally
taking $\epsilon\rightarrow0$ and relabelling $\bar{t} \to t$,
$\rho \to r$, we obtain \ba
ds^2&=&\frac{r-b}{r_0}dt^2-\frac{r_0}{r-b} dr^2-r_0 r d\Omega^2,
\label{gstatic}\\
\e^{2\phi}&=&\frac{r}{r_0}, \qquad
F=\frac{1}{\sqrt{2}r_0} dr\wedge dt,\label{phiFstatic}
\ea which is a generalisation of (\ref{gb0})-(\ref{phiFb0}). This
configuration, which was previously given in \cite{GS} (the
``horizon plus infinite throat case'') is an exact solution of
EMDA.

The Penrose diagrams corresponding to the different values of $b$
($b<0$, $b=0$ and $b>0$) are shown in Fig.1. For $b>0$,
(\ref{gstatic}) describes a static black hole, with a spacelike
singularity $r=0$ hidden behind a horizon located at $r=b$. The
corresponding Penrose diagram is identical to that of the
Schwarzschild black hole. However the spacetime is not
asymptotically flat. The case $b=0$ corresponds to the extreme
black hole (\ref{gb0}), with a null singularity. Finally for
$b<0$, (\ref{gstatic}) describes a naked timelike singularity located at
$r=0$.

The two parameters $r_0$ and $b$ have somewhat different physical
meanings. The first describes the electric charge $Q$ of 
the solution which is given
by the flux through a 2-sphere
\begin{equation}\lb{Q}
Q=\frac1{4\pi}\int\e^{-2\phi} F^{0r}\sqrt{g}\,d\Omega=\frac{r_0}{\sqrt{2}}.
\end{equation}
It is associated, not with a specific black hole, but rather
with a given linear dilaton background and its black hole family. The
parameter $b$ characterizing a black hole is proportional to its mass,
as suggested by the following thermodynamical argument. Carrying
out the Wick rotation $\tau=i t$ and putting $r=b+x^2$ the two
dimensional sector ($t,r$) of (\ref{gstatic}) becomes :
\begin{equation}
ds^2= \frac{x^2}{r_0} d\tau^2+4 r_0 dx^2,
\end{equation}
which is free from conical singularity provided $\tau$ is 
an angular variable of period
$P=4 \pi r_0$. Therefore the Hawking temperature of the black hole is
\begin{equation}
T=\frac{1}{P}=\frac{1}{4 \pi
r_0}
\end{equation}
(computation of the surface gravity gives the same value).
Assuming that the entropy is given as usual by the quarter of the
horizon area
\begin{equation}
S=\frac{A}{4}=\pi r_0 b\,,
\end{equation}
we obtain from the thermodynamic first law
\begin{equation}\lb{Mtherm}
d{\cal M} = T dS
\end{equation}
the value
\begin{equation}\lb{massnaive} 
{\cal M} =  \frac{b}{4}.
\end{equation}
This  mass value will be confirmed in Sect. 5 by an independent
computation using the method of Brown and York \cite{BY}.

Note that here the parameter $r_0$ is kept fixed, being related to the linear 
dilaton background, not to the hole. It would be interesting, however, to 
check whether this is correct in the framework of the
grand canonical ensemble approach
\cite{GiHa77,BR}. This is a non-trivial task because of the lack of
asymptotic flatness (in particular, the value of the Maxwell potential 
entering the standard first law for charged black holes diverges at infinity),
and remains beyond the scope of the present paper, where we rather use
microcanonical considerations.   

Now we give the $\sigma$-model representation of the solutions
(\ref{gstatic})-(\ref{phiFstatic}). In the stationary sector of
EMDA, reduction to three dimensions can be performed by using the
metric ansatz
\begin{equation}
g_{\mu\nu}=\left(
\begin{array}{cc}
f&-f\omega_i\\
-f \omega_i&-f^{-1}h_{ij}+f\omega_i\omega_j
\end{array}
\right)\,,
\end{equation}
and parametrizing the vector field $A_\mu$ by an electric
potential $v$ and a magnetic potential $u$ according to
\begin{equation}
F_{i0}=\frac{1}{\sqrt{2}}\partial_i v, \quad
e^{-2\phi}F^{ij}+\kappa {\tilde F}^{ij} =\frac{f}{\sqrt{2
h}}\epsilon^{ijk}\partial_k u.
\end{equation}
From the mixed components of the four--dimensional Einstein
equations, $\omega_k$ may be dualized to a twist potential $\chi$
by \cite{GaKe94}
\begin{equation}
\partial_i\chi +v\partial_iu-u\partial_iv = -
\frac{f^2}{\sqrt{h}}h_{ij} \epsilon^{jkl}\partial_k\omega_l.
\end{equation}
The six potentials $f$, $\chi$, $u$, $v$, $\phi$, $\kappa$
parametrize a target space isomorphic to the coset $Sp(4,R)/U(2)$
\cite{Ga95}, the EMDA field equations reducing to those of the
corresponding three-dimensional $\sigma$ model. A matrix
representative of this coset can be chosen to be the symmetric
symplectic matrix \cite{GaKe95}
\begin{equation} \label{MPQ}
M=\left(\begin{array}{crc}
P^{-1}&P^{-1}Q\\
QP^{-1}&P+QP^{-1}Q\\
\end{array}\right),
\end{equation}
where $P$ and $Q$ are the real symmetric $2 \times 2$ matrices
\begin{equation}
P=-{\rm e}^{-2\phi} \begin{pmatrix}  v^2-f{\rm e}^{2\phi} & v \\
v & 1 \end{pmatrix}, \quad Q= \begin{pmatrix} vw-\chi & w \\ w &
-\kappa \end{pmatrix},
\end{equation}
with $w=u-\kappa v$.

When all the potentials depend on a single scalar potential
$\sigma$ (such as in the case of spherical symmetry), this
potential can always be chosen to be harmonic ($\nabla_h^2\sigma=
0$). The remaining $\sigma$-model equations, which reduce to the
geodesic equation on the target space, with $\sigma$ as affine
parameter, are solved by \cite{ClGa96}
\begin{equation}\label{geosig}
M=A \e^{B\sigma},
\end{equation}
where the matrices $A$ and $B$ are constant. In the case of
asymptotically flat solutions, the vacuum matrix $A$ is
$diag(+1,-1,+1,-1)$ in the gauge $\sigma(\infty) = 0$. In the
present case, the matrix M associated with the linear dilaton
black hole solution (\ref{gstatic})-(\ref{phiFstatic}) is
\begin{equation}
M_{\ell}=\left(
\begin{array}{cccc}
\frac{r_0}{r-b}&-\frac{r}{r-b}&0&0\\
-\frac{r}{r-b}&\frac{b}{r_0}\frac{r}{r-b}&0&0\\
0&0&-\frac{b}{r_0}&-1\\
0&0&-1&-\frac{r_0}{r}
\end{array}
\right)\,. \label{mstatic}
\end{equation}
The corresponding harmonic potential is
\begin{equation}
\sigma=-\frac{1}{b}\ln\left|\frac{r-b}{r}\right| \,,
\end{equation}
and the matrices $A$ and $B$ are
\begin{equation}
A=\left(
\begin{array}{cccc}
0&-1&0&0\\
-1&\frac{b}{r_0}&0&0\\
0&0&-\frac{b}{r_0}&-1\\
0&0&-1&0
\end{array}
\right), \quad B=\left(
\begin{array}{cccc}
0&0&0&0\\
-r_0&b&0&0\\
0&0&0&-r_0\\
0&0&0&-b
\end{array}
\right) \label{AB}
\end{equation}

In the case $Tr(B^2)=0$,  
(\ref{geosig}) is a null geodesic in the target space \cite{ClGa96},
and the corresponding solution is of the Majumdar-Papetrou type.
Here this occurs for a vanishing black hole mass $b=0$. 
As obvious from (\ref{gb0}), the spatial metric $h_{ij}$ is flat in
this case, so that one can linearly superpose harmonic potentials $\sigma_i
= 1/r_i$ to construct multi-center solutions
\begin{equation}
ds^2=\frac{1}{\ds \sum_i \sigma_i} dt^2-\sum_i \sigma_i d\vec{x}^2.
\end{equation}
This is related to the fact that the linear dilaton vacuum
solution (\ref{gb0})-(\ref{phiFb0}) is BPS, in the sense that it
has $N=2$ supersymmetry \cite{KP}, i.e. half of the original
symmetries of $N=4$, $D=4$ supergravity are preserved.

\section{Kerr extension}
\setcounter{equation}{0}

In this section we wish to generalize the static solution
(\ref{gstatic})-(\ref{phiFstatic}) to a rotating  one.  To do
this, we shall first find the linear transformation $U$
\begin{equation}
M\rightarrow U^T M U \label{translaw}
\end{equation}
that transforms the matrix $M_S$ associated with the Schwarzschild
solution into the matrix $M_{\ell}$ associated with the static
black hole solution (\ref{mstatic}). Note that this
transformation changes the matrix representative, i.e. the
potentials contained in the matrices $P$  and $Q$, without
changing the reduced three-space metric $h_{ij}$, which is
invariant under $U$. Then we shall apply the same transformation
to the matrix representative $M_K$ of the Kerr solution to
generate a rotating black hole metric $M_{\ell K}$.

The Schwarzschild solution \ba ds^2 &=& (1-\frac{2M}r)dt^2 -
(1-\frac{2M}r)^{-1}dr^2 - r^2 d\Omega^2, \\ e^{2\phi} &=& 1, \quad
\kappa = 0, \quad F = 0 \ea has the same reduced 3-metric as
(\ref{gstatic}) provided we identify \be b = 2M.  \end{equation} The
corresponding Schwarzschild matrix is
\begin{equation}
M_S=\left(
\begin{array}{cccc}
\frac{r}{r-2M}&0&0&0\\ 0&-1&0&0\\ 0&0&\frac{r-2M}{r}&0\\ 0&0&0&-1
\end{array}
\right)\,. \label{ms}
\end{equation}
Using (\ref{translaw}), (\ref{ms}) and (\ref{mstatic}) we derive
the expression of the  matrix $U$
\begin{equation}\label{U}
U=\left(
\begin{array}{cccc}
-\sqrt{\frac{r_0}{2M}}&\sqrt{\frac{2M}{r_0}}&0&0\\
\sqrt{\frac{r_0}{2M}}&0&0&0\\ 0&0&0&\sqrt{\frac{r_0}{2M}}\\
0&0&\sqrt{\frac{2M}{r_0}}&\sqrt{\frac{r_0}{2M}}
\end{array}
\right)\,.
\end{equation}

Consider now the Kerr-NUT solution \ba ds^2 & = &
\frac{\Delta-a^2\sin^2\theta}{\Sigma}(dt-\omega d\varphi)^2 -
\Sigma\bigg(\frac{dr^2}{\Delta}+d\theta^2+\frac{\Delta\sin^2\theta}
{\Delta-a^2\sin^2\theta}\,d\varphi^2 \bigg), \\ e^{2\phi} &=&1,
\quad \kappa = 0, \quad F = 0, \ea with \ba \Delta & = & r^2-2 M
r+a^2-N^2, \quad  \Sigma =  r^2+(N+a\cos\theta)^2, \\ \omega & = &
-2\,\frac{N\Delta\cos\theta+a(Mr+N^2)\sin^2\theta} {\Delta -
a^2\sin^2\theta}.   \ea The corresponding matrix $M_K$ is
\begin{equation}
M_{K}=\left(
\begin{array}{cccc}
f^{-1}&0&-\chi f^{-1}&0\\ 0&-1&0&0\\ -\chi f^{-1}&0&f+\chi^2
f^{-1}&0\\ 0&0&0&-1\\
\end{array}
\right) , \label{mk}
\end{equation}
where \be f=\frac{\Delta-a^2\sin^2\theta}{\Sigma}, \quad
\chi=-2\frac{N(M-r)+aM\cos\theta}{\Sigma}\,.  \end{equation} Applying the
transformation (\ref{translaw}) to the matrix $M_K$ we obtain
\begin{equation} \label{mstationary}
M_{\ell K}=U^TM_{K} U=\left(
\begin{array}{cccc}
\frac{r_0(1-f)}{2M f}&-\frac{1}{f}&0&\frac{r_0\chi}{2M f}\\
-\frac{1}{f}&\frac{2M}{r_0 f}&0&-\frac{\chi}{f}\\
0&0&-\frac{2M}{r_0}&-1\\ \frac{r_0\chi}{2M
f}&-\frac{\chi}{f}&-1&\frac{r_0}{2Mf}(\chi^2+f^2-f)\\
\end{array}
\right)\,.
\end{equation}
Comparing (\ref{mstationary}) and (\ref{MPQ}) we obtain the
rotating black hole solution generalizing
(\ref{gstatic})-(\ref{phiFstatic}) \ba ds^2 & = &
\frac{\Delta-a^2\sin^2\theta}{\Gamma}(dt-\omega d\varphi)^2 -
\Gamma\bigg(\frac{dr^2}{\Delta}+d\theta^2+\frac{\Delta\sin^2\theta}
{\Delta-a^2\sin^2\theta}\,d\varphi^2 \bigg), \label{nkg}\\
\e^{2\phi} & = & \frac{r^2+(N+a\cos\theta)^2}{\Gamma}, \qquad
\kappa = \frac{r_0}{M}\,\frac{N(r-M) -
aM\cos\theta}{r^2+(N+a\cos\theta)^2}, \label{nkda}\\  v & = &
\frac{r^2+(N+a\cos\theta)^2}{\Gamma},\qquad
u=\frac{r_0}{M}\,\frac{N(r-M)-aM\cos \theta}{\Gamma}, \label{nkvu}
\ea with \ba \Delta & = & r^2-2 M r+a^2-N^2, \quad  \Gamma =
\frac{r_0}{M}(Mr+N^2+aN\cos\theta), \\ \omega & = &
-\frac{r_0}{M}\,\frac{N\Delta\cos\theta+a(Mr+N^2)\sin^2\theta}
{\Delta - a^2\sin^2\theta}.   \ea

This new solution can again be slightly generalized by the action
of the S-duality transformation $SL(2,R)$
\begin{equation}
\tilde{z}=\tilde{\kappa}+i \e^{-2\tilde{\phi}}=\frac{\alpha
z+\beta} {\gamma z+\delta}, \quad \tilde{v}=\delta v+\gamma u,
\quad \tilde{u}=\beta v+\alpha u \qquad (\alpha\delta-\beta\gamma = 1),
\end{equation}
which generates a magnetic charge without modifying the spacetime metric. 

We now focus on the case $N=0$. The stationary solution
(\ref{nkg})-(\ref{nkvu})  is then significantly simplified to
\ba
ds^2 & = & \frac{r^2-2Mr+a^2}{r_0 r}dt^2-r_0
r\left[\frac{dr^2}{r^2-2Mr+a^2}
+d\theta^2+\sin^2\theta\bigg(d\varphi-\frac{a}{r_0 r}dt
\bigg)^2\right]\,, \label{nkgN0} \\
F & = &
\frac1{\sqrt{2}}\bigg[\frac{r^2-a^2\cos^2\theta}{r_0r^2}\,dr\wedge dt
+ a\sin 2\theta\,d\theta\wedge\bigg(d\varphi-\frac{a}{r_0 r}dt
\bigg)\bigg]\,, \lb{nkFN0} \\
e^{-2\phi} & = & \frac{r_0r}{r^2+a^2\cos^2\theta}\,, \qquad \kappa =
-\frac{r_0a\cos\theta}{r^2+a^2\cos^2\theta}\,.\lb{nkdaN0}
\ea
The Maxwell two-form is derivable from the following four-potential
\begin{equation}\lb{pots}
A=\frac1{\sqrt{2}}\left(\frac{r^2+a^2\cos^2\theta}{r_0 r}dt+
a\sin^2\theta\,d\varphi\right),
\end{equation}
where the gauge is chosen such that the vector magnetic potential be
regular on the axis $\theta = 0$.

Although the metric (\ref{nkgN0}) was derived from the Kerr metric, it differs
in its asymptotic behavior ($r \to \infty$), which is the same as
for the linear dilaton metric (\ref{gb0}), and in its behavior
near $r = 0$. In the case of the Kerr metric, $r = 0$ is the
equation of a disk through which the metric can be continued to
negative $r$, while in (\ref{nkgN0}) $r = 0$ is a timelike line
singularity. It follows that the Penrose diagrams of
(\ref{nkgN0}) for the three cases  $M^2>a^2$, $M^2=a^2$ and
$M^2<a^2$ are identical, not to those of the Kerr spacetime, but
rather to those of the Reissner-Nordstr\"{o}m spacetime, with the
charge replaced by the angular momentum parameter $a$.

The massless case $M=0$ is of particular interest. The Kerr metric
with $M = 0$ representing flat Lorentzian spacetime, it follows
that the reduced 3-metric $h_{ij}$ in (\ref{nkgN0}), which is
identical to that of the Kerr metric with $M = 0$, represents
flat Euclidean space. This property is characteristic of the
generalized Israel-Wilson-Perj\`es (IWP) solutions. Indeed,
following the methods of \cite{ClGa96}, we can show that for $M =
0$ the matrix $M_{\ell K}$ depends on the two harmonic potentials
\begin{equation}
\sigma=\frac{r}{r^2+a^2 \cos^2\theta}\,, \quad \tau=
\frac{\cos\theta}{r^2+a^2\cos^2\theta},
\end{equation}
through
\begin{equation}
M_{\ell K}=A\,\e^{B\sigma+C\tau}, \label{m0}
\end{equation}
where the matrices $A$ and $B$ are those of (\ref{AB}) with $b =
2M = 0$, and
\begin{equation}
C=\left(
\begin{array}{cccc}
0&0&0&0\\ 0&0&0&a\, r_0\\ a \,r_0&0&0&0\\ 0&0&0&0
\end{array}
\right).
\end{equation}
These matrices are such that $B^2=C^2=BC=CB=0$, so that the matrix
$M_{\ell K}$ is linear in the two harmonic potentials,
\begin{equation}
M_{\ell K}=A(I+B\sigma+C\tau)\,.
\end{equation}
It follows that given any two harmonic potentials $\sigma$,
$\tau$, we can construct the IWP generalization of the rotating
black hole solution (\ref{nkg})-(\ref{nkvu}) with $M = N = 0$  \ba
ds^2&=&\frac{1}{r_0\sigma}(dt-\omega_i\, dx^i)^2-r_0\sigma
d^3x,\label{iwpg}\\ A  &=& \frac1{r_0\sigma}(dt-\omega_i\, dx^i),
\label{iwpA}\\ \e^{-2 \phi}&=&r_0\sigma\, ,\quad\kappa=-a
r_0 \tau\, , \label{iwpz} \ea with
\begin{equation}
\partial_i\omega_j=\frac{ar_o}2\epsilon_{ijk}\partial^k\tau\,.
\end{equation}

The general IWP solution (\ref{iwpg})-(\ref{iwpz}) was previously
constructed in \cite{Ka94}. The multi-center solutions given in
\cite{Ka94}, \cite{ClGa96} are asymptotically flat, and so do not
explicitly include the solution (\ref{nkg})-(\ref{nkvu})  with $M
= N = 0$. This solution can be recovered from the one-particle
rotating IWP solution given in \cite{Ka94} by taking the infinite
mass limit. It was shown in \cite{Ka94} that the general IWP
solution is BPS, i.e. has N = 2 supersymmetry when embedded in N
= 4, D = 4 supergravity, and this result applies also to our new
non-asymptotically flat solutions  (\ref{nkg})-(\ref{nkvu})  with
$M = N = 0$.

\section{ Six-dimensional vacuum dual}
\setcounter{equation}{0}

As shown in \cite{bremda}, the four-dimensional EMDA theory is
equivalent to a sector of sourceless six-dimensional general
relativity with two Killing vectors. That is, any solution
$(ds_4^2, \; A_\mu, \; \phi, \; \kappa$) of EMDA may be lifted to
the Ricci-flat six-dimensional metric \begin{equation}\label{red64a} ds_6^2 =
ds_4^2 - \e^{-2\phi}\,\xi^2 - \e^{2\phi}\,(\zeta +
\kappa\xi)^2\,, \end{equation} with \ba\label{red64b} \xi \equiv d\chi +
\sqrt2A_{\mu}\,dx^{\mu}\,, \qquad
\zeta \equiv d\eta + \sqrt2B_{\mu}\,dx^{\mu}\,,   \\ \notag\\
F_{\mu\nu}(B) \equiv \e^{-2\phi}\tilde{F}_{\mu\nu}(A) - \kappa
F_{\mu\nu}(A)\,. \ea Note that the $S$-duality  of the
four-dimensional dilaton-axion theory is here explicitly realized
as the group of linear transformations of the 2-plane
($\chi,\,\eta$).

In the present case we obtain from Eqs. (\ref{nkg})-(\ref{nkvu})
\ba A & = & \frac{1}{\sqrt2\Gamma}\left(
[r^2+(N+a\cos\theta)^2]\,dt + \frac{r_0}{M}[N\Delta\cos\theta + 
a(Mr+N^2)\sin^2\theta]\,d\varphi \right),  \\ B & = &
\frac{(M^2+N^2)r_0}{\sqrt2M^2\Gamma}\bigg( (N+a\cos\theta)\,dt  -
r_0r\cos\theta\,d\varphi \bigg), \ea leading for the
six-dimensional metric to a long and unenlightening expression.
In the case $N = 0$, after rescaling $\eta \to r_0\eta$ this
reduces to 
\ba\label{n6N0} ds_6^2 & = & -\frac{2M}{r_0}\,dt^2 -
2\,d\chi\,dt   - \frac{r_0}{r}(d\chi+a\,d\varphi-a\cos\theta\,d\eta)^2 
\nonumber \\ & & - \frac{r_0r}{\Delta}\,dr^2 -
4r_0r\,d\Omega_3^2.   \ea In (\ref{n6N0}),   \be d\Omega_3^2 =
\frac14\left(d\theta^2 + \sin^2\theta\,d\varphi^2  + (d\eta -
\cos\theta\,d\varphi)^2\right) \end{equation} is the metric of the
three-sphere if $0 < \theta < \pi$, and the angles $\eta$ and
$\varphi$ are defined modulo $2\pi$.  This can be seen from the
fact that the coordinate transformation \begin{equation} x+iy =
r\cos(\theta/2)\e^{i\frac{\eta-\varphi}2}, \quad z+iw =
r\sin(\theta/2)\e^{i\frac{\eta+\varphi}2}, \end{equation} describes the
embedding of the three-sphere $x^2+y^2+z^2+w^2=r^2$ in
four-dimensional Euclidean space \be dx^2+dy^2+dz^2+dw^2=dr^2 +
\frac{r^2}4\left(d\theta^2 + d\eta^2 + d\varphi^2  -
2\cos\theta\,d\eta\,d\varphi\right).  \end{equation}

Remarkably, the metric (\ref{n6N0}) is asymptotically flat.
Defining new coordinates by 
\ba\label{new6} \rho & = & 2\sqrt{r_0r}, \nonumber \\
\bar{\theta} & = & \theta/2, \qquad \varphi_{\pm} = (\varphi\pm
\eta)/2, \nonumber \\ d\psi & = & (2M/r_0)^{1/2}(dt +
(r_0/2M)\,d\chi), \\ d\tau & = &
(2M/r_0)^{-1/2}\,d\chi, \nonumber \ea it can be put
in the form of a direct product \be ds_6^2 = - d\psi^2 + ds_5^2,
\end{equation} where \ba\label{mype} ds_5^2 & = & d\tau^2 -
\frac{\mu}{\rho^2}\bigg(d\tau + (\bar{a}/2)(d\varphi -
\cos\theta\,d\eta)\bigg)^2 -
\frac{d\rho^2}{1-\mu\rho^{-2}+\mu\bar{a}^2\rho^{-4}} -
\rho^2\,d\Omega_3^2  \nonumber \\ & = & d\tau^2 -
\frac{\mu}{\rho^2}\bigg(d\tau +
\bar{a}\sin^2\bar{\theta}d\varphi_+ +
\bar{a}\cos^2\bar{\theta}d\varphi_-\bigg)^2 \\ & & -
\frac{d\rho^2}{1-\mu\rho^{-2}+\mu\bar{a}^2\rho^{-4}} -
\rho^2\bigg(d\bar{\theta}^2 + \sin^2\bar{\theta}d\varphi_+^2 +
\cos^2\bar{\theta}d\varphi_-^2\bigg) \nonumber \ea ($\mu = 8Mr_0$,
$\bar{a} = (2r_0/M)^{1/2}a$) is the five-dimensional Myers-Perry
black hole \cite{MP5} with two equal angular momentum parameters.
Accordingly, the six-dimensional metric (\ref{n6N0}) enjoys the
symmetry group $SO(3) \times U(1) \times U(1) \times U(1)$
generated by the six Killing vectors \ba & L_1 =
\cos\eta\,\p{\theta} -
\frac{\ds\sin\eta}{\ds\sin\theta}\,\p{\varphi} -
\cot\theta\,\sin\eta\,\p{\eta}, \qquad & L_4 = \p{\varphi},
\nonumber\\ & L_2 = \sin\eta\,\p{\theta} +
\frac{\ds\cos\eta}{\ds\sin\theta}\,\p{\varphi} +
\cot\theta\,\cos\eta\,\p{\eta}, \qquad & L_5 = \p{\tau}, \\ & L_3
= \p{\eta}, \qquad\qquad\qquad\qquad\qquad\qquad\qquad\;\;\;\;\;
& L_6 = \p{\psi}. \nonumber \ea It is interesting to observe that the
four-dimensional rotating solution did not possess the spherical
symmetry, the origin of the $SO(3)$ component in six dimensions may be
traced to the
$Sp(4,R)\sim SO(3,2)$ invariance of stationary EMDA configurations. 

Several particular cases are interesting. The Myers-Perry black
hole is extremal for $4\bar{a}^2 = \mu$ ($a^2 =M^2$). For
$\bar{a} = 0$ ($a = 0$), the spacetime (\ref{mype}) reduces to
the static black hole \be ds_5^2 = (1-\frac{\mu}{\rho^2})d\tau^2
- \frac{d\rho^2}{\ds 1-\frac{\mu}{\rho^2}} - \rho^2\,d\Omega_3^2,
\end{equation} leading for the six-dimensional metric to the higher symmetry
group $SO(4) \times U(1) \times U(1)$.  ( Note that $SO(4) \sim SO(3)
\times SO(3)$, where the first $SO(3)$ is the spatial spherical
symmetry, and the second is the internal $SO(3)$ subgroup of
$SO(3,2)$).  For $\mu = 0$ (\ref{mype}) reduces to
five-dimensional Minkowski space.

For $M= 0$ the coordinate transformation (\ref{new6}) breaks
down. In this case (\ref{n6N0}) reduces to the six-dimensional
asymptotically flat metric 
\begin{equation} ds_6^2 = -2\,dt\,d\chi -
\frac{4r_0^2}{\rho^2}\bigg(d\chi + a(d\varphi -
\cos\theta\,d\eta)\bigg)^2 -
\frac{d\rho^2}{1+16r_0^2a^2\rho^{-4}} - \rho^2\,d\Omega_3^2.  \end{equation}
Finally, for $M = 0$ and $a = 0$, this reduces to the
six-dimensional metric \begin{equation} ds_6^2 = -2\,dt\,d\chi -
\frac{4r_0^2}{\rho^{2}}\,d\chi^2 - d\rho^2 -
\rho^2\,d\Omega_3^2.  \end{equation} This has flat spatial four-sections,
and is easily seen to be a special case of the six-dimensional
multi-center metric, generalizing the five-dimensional
``antigravitating'' solutions given by Gibbons \cite{gibbons82}
(Eq. (18), see also \cite{spat}, Eq. (40))   \begin{equation} ds_6^2 =
2\,dt\,d\psi - F\,d\psi^2 - d{\bf x}^2, \end{equation} with $F({\bf x})$ an
arbitrary harmonic function of the four spatial dimensions.

\section{Mass, spin and the first law}
\setcounter{equation}{0}

Let us discuss the physical properties of the rotating dilaton black
hole solution with $N = 0$ (\ref{nkgN0}).
The space-time contains an event horizon
located at $r=r_+$ which is the largest root of the equation
$\Delta=(r-r_-)(r-r_+)=0$:
\begin{equation} \lb{eh}
r_{\pm}=M\pm\sqrt{M^2-a^2}.
\end{equation}
This expression is the same as for the Kerr metric, but one has to
keep in mind that $M$ is no longer the physical mass of the solution.

Again as the Kerr space-time, the space-time (\ref{nkgN0}) does not
admit a globally timelike Killing vector field, and therefore
contains an ergosphere  whose boundary
corresponds to vanishing of the norm of the Killing vector
$\partial_t$ (i.e.  is given by the largest root of $g_{tt}=0$):
\begin{equation} \lb{erg}
r_e=M +\sqrt{M^2-a^2\cos^2\theta}.
\end{equation}
All physical frames inside the ergosphere are rotating. The angular
velocity of the horizon is determined by vanishing of the norm of the
linear combination of the timelike and azimuthal Killing vectors
$\partial_{t}+\Omega_h\partial_{\varphi}$ on the horizon (\ref{eh})
\begin{equation}\lb{omh}
\Omega_h=\frac{a}{r_0r_+}.
\end{equation}

Let us compute the mass and the angular momentum of the NUT-less
stationary black hole (\ref{nkgN0}) using the method developed by
Brown and York \cite{BY}. Consider a spacetime region $M$ foliated by
a family $\Sigma$ of spacelike slices of two-boundary $B$, which we
shall assume to be 2-spheres $r =$ constant. The metric on $M$
is written in the ADM form \cite{ADM} as
\be
ds^2 = -N^2\,dt^2 + h_{ij}(dx^i + V^i\,dt)(dx^j + V^j\,dt).
\end{equation}
From (\ref{nkgN0}) the unit normal to $\Sigma$ is $u_{\mu} =
-N\delta^0_{\mu}$, with $N = \sqrt{\Delta/r_0r}\,$, and the induced
metric $h_{\mu\nu} = g_{\mu\nu} + u_{\mu}u_{\nu}$ on $\Sigma$ is
\be
h_{ij}dx^idx^j = r_0r\,\left(\frac{dr^2}{\Delta} + d\Omega^2\right).
\end{equation}
The timelike component ${}^3B$ of the three-boundary of $M$ is the
product of $B$ with segments of timelike lines orthogonal to $\Sigma$
at $B$. The unit normal to ${}^3B$ is $n_{\mu} =
\sqrt{r_0r/\Delta}\,\delta^r_{\mu}$, and the induced metric
$\sigma_{\mu\nu} = h_{\mu\nu} - n_{\mu}n_{\nu}$ on $B$ is
$\sigma_{ab}dx^adx^b = r_0r\,d\Omega^2$.

The surface stress-energy-momentum tensor $\tau^{ij}$ is defined by
the functional derivative of the action with respect to the
three-metric on ${}^3B$. The normal and tangential projections of
$\tau^{ij}$ on $B$ lead to the proper energy surface density and
proper momentum surface density
\ba
\varepsilon  & \equiv & u_iu_j\tau^{ij} =
-\frac{1}{\sqrt{\sigma}}\frac{\delta \tilde{S}}{\delta N}, \\ j_a  &
\equiv & -\sigma_{ai}u_j\tau^{ij} =
\frac{1}{\sqrt{\sigma}}\frac{\delta \tilde{S}}{\delta V^a}.
\ea
Here the ``renormalized'' action $\tilde{S}$ is the sum
\be
\tilde{S} = S_{(g)} + S_{(m)} - S_{(0)},
\end{equation}
where $S_{(g)}$ and $S_{(m)}$ are the ``gravitational'' and ``matter''
pieces of the action (\ref{ac}), evaluated at a given classical
solution, and $S_{(0)}$ is the value of the action $S$ at a reference
or background ``vacuum'' classical solution. Such a substraction is
necessary in the case of a non-asymptotically flat metric \cite{BY}
\cite{HaHo95}. Here the natural choice for the vacuum in a given
charge sector is the linear dilaton metric (\ref{gb0}), corresponding
to $M=a=0$ and the specified value $Q = r_0/\sqrt{2}$ of the electric
charge.

We first discuss the computation of the proper energy surface density
$\varepsilon$. The gravitational contribution is
\begin{equation}\lb{Egrav}
\varepsilon_{(g)}=\frac{1}{8\pi}k,
\end{equation}
where $k$ is the trace of the extrinsic curvature of $B$ as embedded
in $\Sigma$,
\be
k = -\sqrt{\frac{\Delta}{r_0r}}\,\frac1r.
\end{equation}
The matter contribution originates solely from the gauge field piece
of the action (\ref{ac}) and is given by
\be\lb{Ematt}
\varepsilon_{(m)}= \frac1{N\sqrt{\sigma}}A_0 \Pi^r,
\end{equation}
Here $\Pi^i = N\sqrt{h} E^i/4\pi$ is the canonical momentum conjugate to
$A_i$, the radial component $E^r$ reads
\begin{equation} \lb{Er}
E^r=\e^{-2\phi} F^{ rt} +\kappa {\tilde F}^{rt}=-\frac1{\sqrt{2} r}
\end{equation}
(it follows from (\ref{Er}) that the electric charge is given as in
the static case by $Q = -4\pi\Pi^r = r_0/\sqrt{2}$), so that we obtain from
(\ref{pots}) 
\be
\varepsilon_{(m)} = -\frac1{8\pi
r_0}\sqrt{\frac{r_0r}{\Delta}}\frac{r^2+a^2\cos^2\theta}{r^2}.
\end{equation}

It is worth noting that in the Brown-York formalism there are two
quantities associated with energy: the proper energy density
$\varepsilon$, whose integral gives the Brown-York total quasilocal
energy
$E = \int_B d^2x \sqrt{\sigma}\varepsilon$, and the mass 
$\cal M$, which is given by the following integral over a 
large sphere at spatial infinity
\begin{equation}\lb{massBY}
{\cal M}=-\int_B d^2x \sqrt{\sigma}u_0\varepsilon = -\int_B d^2x
\sqrt{\sigma}u_0(\varepsilon_{(g)}+\varepsilon_{(m)}-\varepsilon_{(0)}).
\end{equation}
Both quantities give the same answer for asymptotically flat spacetimes,
but not for non-asymptotically flat ones. This difference, in particular,
manifests itself for asymptotically AdS black holes, and it was found
that it is the mass $\cal M$ which gives the correct form of the
first law, while the energy gives the redshifted quantity \cite{BrCrMa94}. 
So here we will calculate the Brown-York mass (\ref{massBY}). 
For asymptotically flat black holes, such as the Kerr-Newman one, the
second term is zero and is usually ignored. In our case it is
non-zero, and indeed linearly divergent, but is exactly cancelled by the
electromagnetic contribution of the linear dilaton background
evaluated with the same boundary data for the fields $\sigma$, $N$ and
$A_0$ \cite{HaRo95}, and in the present case with the same value of
the electric charge. The linear divergence of
the first term is cancelled by substracting the gravitational
contribution of the linear dilaton background
\be
\varepsilon_{(g0)} = - \frac1{8\pi\sqrt{r_0r}},
\end{equation}
leading to the finite result
\be\lb{finalmass}
{\cal M} = \frac{M}2 = \frac{b}4\,,
\end{equation}
in accordance with the value (\ref{Mtherm}) for the static case (we
recall that the parameter $M$ is the mass of the Kerr black hole
which is transformed into the black hole metric (\ref{nkgN0}) by the
group transformation (\ref{translaw})).
Here (as usual) the black hole mass is given by the finite part of
the gravitational contribution, the (divergent) matter contribution
which is independent of the black hole parameters $M$ and $a$ being
completely compensated by the background substraction. 

The angular momentum is computed along similar lines. The
gravitational and matter contributions to the azimuthal proper
momentum surface density are
\begin{equation}
j_{(g)\varphi} = -\frac1{8\pi}n_lK_{\varphi}^l\,, \quad j_{(m)\varphi}
= -\frac1{\sqrt{\sigma}}A_{\varphi}\Pi^r,
\end{equation}
where $K_{ij}$ is the extrinsic curvature of $\Sigma$. The total 
angular momentum is given by the integral
\be
J=\int_B d^2x\sqrt{\sigma}j_{\varphi}=\int_B
d^2x\sqrt{\sigma}(j_{(g)\varphi}+j_{(m)\varphi}) \label{MBY}
\end{equation}
(the spherically symmetric background does not contribute to
(\ref{MBY})). Again the matter contribution $j_{(m)\varphi}$, which
vanishes in the Kerr-Newman case, is often ignored (see however
\cite{Wa93} and references therein). In the case of our rotating
black hole metric (\ref{nkgN0}) we obtain $j_{(g)\varphi} =
a\sin^2\theta/16\pi r$, $j_{(m)\varphi} = 2j_{(g)\varphi}$, leading
to
\begin{equation}\lb{Ja}
J=\frac{a r_0}{2}.
\end{equation}

Now we are in a position to check that our black holes obey the
differential first law of black hole mechanics:
\begin{equation}\lb{dfl}
d{\cal M} =TdS+\Omega_h dJ.
\end{equation}
The Hawking temperature of rotating dilaton black holes can be
computed either via the surface gravity, or using euclidean
continuation, both leading to
\begin{equation} \lb{T}
T=\frac{r_+-M}{2\pi r_0 r_+}.
\end{equation}
The entropy is given by the quarter of the horizon area
\begin{equation} \lb{S}
S=\pi r_0 r_+ .
\end{equation}
Using (\ref{finalmass}), (\ref{T}), (\ref{S}), (\ref{omh}) and
(\ref{Ja}) it is straightforward to verify the first law (\ref{dfl}),
in which differentiations are performed for a fixed value of the
electric charge $r_0/\sqrt{2}$ which is characteristic of the linear
dilaton background.

\section{Geodesics and test scalar field}
\setcounter{equation}{0}

To get further insight into the nature of our solutions, consider
first geodesics. The Hamilton-Jacobi equation is separable in the
general rotating linear dilaton black hole space-time (\ref{nkgN0})
in exactly the same way as as in the usual Kerr metric. This
indicates that a Stachel-Killing tensor exists in the present case
too. The variables in the equation \be \frac{\partial S}{\partial
x^\mu}\frac{\partial S}{\partial x^\nu} g^{\mu\nu}=\mu^2
\end{equation}
are separated by setting
\be
S=-{\cal E} t + L\varphi +\Theta(\theta) + {\cal R} (r),
\end{equation}
where  ${\cal E}$ is the conserved energy  and $L$ is the azimuthal
momentum. We find
\begin{gather}
{{\cal R}'}^2 -\left(\frac{{\cal E} r_0r -aL}{\Delta}\right)^2+
\frac{{\cal K}^2+\mu^2 r_0r}{\Delta}=0, \lb{HJ1}\\
{\Theta'}^2(\theta)+\frac{L^2}{\sin^2\theta}={\cal K}^2. \lb{HJ2}
\end{gather}
The separation constant $\cal K$ is an analogue of Carter's integral
for the Kerr metric, it is equal to the square of the total angular
momentum. Equatorial motion corresponds to ${\cal K}=L$. Finally we
obtain the Hamilton-Jacobi action in the form
\begin{equation}
S=-{\cal E} t + L\varphi \pm \int\sqrt{{\cal
K}^2-\frac{L^2}{\sin^2\theta}} \;d\theta \pm \int\sqrt{({\cal
E}r_0r-aL)^2 - \Delta({\cal K}^2+\mu^2r_0r)}\;\frac{dr}{\Delta}.
\end{equation}
From this one can derive the trajectory and the dependence of
coordinates on time $t$.

Alternatively one may describe timelike and null geodesics  using the
constraint equation  \be \frac{dx^\mu}{d \lambda} \frac{dx^\nu}{d
\lambda} g_{\mu\nu}=\epsilon ,\quad \epsilon=1, 0\end{equation} in
terms of the affine parameter. Reintroducing the (renormalized)
energy and azimuthal momentum as \be {\cal E}=g_{t\mu}\frac{dx^\mu}{d
\lambda} ,\quad L=-g_{\varphi\mu}\frac{dx^\mu}{d \lambda},
\end{equation}
one obtains  the following equation for the equatorial motion:
\be \left(\frac{dr}{d\lambda}\right)^2-\left({\cal
E}-\frac{aL}{r_0r}\right)^2+\frac{\Delta}{(r_0r)^2} (L^2+\epsilon
r_0r)=0,
\end{equation}
which can be rewritten as the conservation equation \be \left(\frac{dr}{d
\lambda}\right)^2 + U_{\rm eff}(r)={\cal E}^2,
\end{equation}
with the effective potential energy \be U_{\rm eff}(r)=
\frac{L^2}{r_0^2}\left(1-\frac{2M}{r}\right) +  \frac{2aL{\cal
E}}{r_0r}+\frac{\epsilon\Delta}{r_0r}.
\end{equation}
For $\epsilon =1$ this potential grows linearly as $r\to\infty$, so
 all timelike  equatorial geodesics are reflected at large $r$. For
null ones the potential is stabilized at infinity, so the geodesics
with ${\cal E}<L/r_0$ are confined, while those with ${\cal E}>L/r_0$
go to infinity at an infinite affine parameter.

Let us discuss the behavior of a minimally coupled test scalar field,
starting with the linear dilaton background $M=a=0$. The Klein-Gordon
equation
\be(\nabla^2+\mu^2)\psi=0 \end{equation}
separates in terms of the spherical harmonics
\be
\psi_{\omega lm}= C_{\omega lm} Y_{lm}(\theta,\varphi) R_{\omega
l}(r) \,\e^{-i\omega t}\,,
\end{equation}
where we assume $\omega>0$, and the radial functions obey the
equation
\be\lb{radial0}
(r^2 R_{\omega l}')'-(\nu^2-1/4+\mu^2r_0r)R_{\omega l}=0,
\end{equation}
where
\be
\nu = \sqrt{(l+1/2)^2-\omega^2r_0^2}.
\end{equation}
For $\mu \neq 0$ the general asymptotic solution of the radial
equation (\ref{radial0}) reads
\be\lb{asmu}
R(r) \sim C_1\,\e^{-2\mu\sqrt{r_0r}} + C_2\,\e^{2\mu\sqrt{r_0r}},
\end{equation}
where one has to put $C_2=0$ for physical solutions, the exact
solution vanishing at infinity being
\be\lb{Kmu}
R(r) = C\,r^{-1/2} K_{2\nu}(2\mu\sqrt{r_0r}),
\end{equation}
with $K_{2\nu}$ a modified Bessel function. These are the massive
modes reflected from infinity whose existence could be guessed from
the above analysis of geodesics.

Near the singularity the mass term can be neglected and we obtain the
following generic solution of the equation (\ref{radial0}) which is
of course exact in the massless case:
\be\lb{asmu0}
R(r) = C_1\,r^{-\nu-1/2} + C_2\,r^{\nu-1/2}.
\end{equation}
For $\omega>(l+1/2)/r_0$ (imaginary $\nu$ so that $\nu=iq$) these two terms
represent ingoing and outgoing waves
\be\lb{asmu0x}
R(r) =\frac{1}{r^{1/2}}\left(C_1\, \e^{-iqx} + C_2\,\e^{iqx}\right),
\end{equation}
where $x=\ln r$. Introducing the current
\be
j^\mu = \frac{i}{2}
\psi^*\overleftrightarrow{\partial^{\mu}}\psi,
\end{equation}
one can see that the flux of the (spherical) mode is
\be
\oint j^r\sqrt{-g} d\Omega =4\pi q (|C_1|^2 -|C_2|^2).\end{equation}
It can be checked that the
solution (\ref{Kmu}) corresponds to $|C_1| =|C_2|$, so it can be
interpreted as initiating from the singularity and reflected back to
it by the linear potential barrier $\mu^2 r/2r_0$.

Now consider our general class of rotating metrics with $N=0$ (\ref{nkgN0}).
The Klein-Gordon equation for the modes
$\psi=\psi(r,\theta)\e^{i(m\varphi-\omega t)}$ takes the form
\begin{equation}
\frac{1}{r_0r}\partial_r\left( \Delta
\partial_r\psi\right)+\frac{1}{r_0r\sin\theta}
\partial_\theta(\sin\theta\partial_\theta\psi)
+\left[\frac{r_0r}{\Delta}\big(\omega-\frac{am}{r_0r}\big)^2
-\frac{m^2}{r_0r\sin^2\theta}-\mu^2\right]\psi= 0\lb{KG}
\end{equation}
and is again separable. In fact, putting $\psi=R(r)\Theta(\theta)$
the above equation is split into the following two ordinary
differential equations
\begin{align}
\partial_r\left( \Delta\partial_r R\right)+ \left(-\mu^2r_0
r+\frac{(r_0r\omega-am)^2}{\Delta}\right)R&={\cal K}^2 R,
\label{radial}\\
-\frac{1}{\sin\theta}\partial_\theta(\sin\theta\partial_\theta\Theta)+
\frac{m^2}{\sin^2\theta}\Theta&={\cal
 K}^2 \Theta \,, \label{angular}
\end{align}
where we keep the same notation for the separation constant as
before. Unlike the Kerr case, now the angular equation
(\ref{angular}) is that for the associated Legendre functions so we
have
\begin{equation}
\begin{array}{lcr}
{\cal K}^2=l(l+1),&l\geq|m|,&l=0,1,2,\ldots
\end{array}
\end{equation}

Let us consider the radial equation (\ref{radial}) in the black hole
case $M>a$ (assuming without loss of generality $a \ge 0$) when a
regular event horizon exists. As usual, the radial wave function
behaves near the horizon as a spherical wave, and the energy spectrum
is continuous. One question to be answered is that of superradiance.
Since the rotating dilaton black hole contains an ergosphere outside
the horizon, it potentially has a superradiant instability. In an
asymptotically flat spacetime such as Kerr, superradiance manifests
itself as a quantum emission of certain modes transporting the
angular momentum of the black hole to infinity. If the asymptotic
region is non-flat, one has to analyse the set of modes more
carefully paying attention to their behavior at infinity
\cite{OtWi00,Wi01}. In particular, the situation is 
different (from that of the Kerr spacetime) in the
AdS-Kerr spacetime, which is neither asymptotically flat nor globally
hyperbolic. As was argued in \cite{Av78} one is free either to
impose reflecting boundary conditions at infinity, or to allow for modes
$\omega>\mu$ to propagate. In the first case the superradiant modes
will be reflected back to the black hole. These waves will generate
stimulated superradiant emission and absorption, whose balance in the 
classical limit corresponds to the classical amplification 
phenomenon \cite{Ga68}. As a result, the superradiant  modes  will
grow exponentially causing  transport of the angular
momentum to the field cloud rotating outside the horizon
\cite{HaRe99}, until the black hole angular momentum is lost
completely. From the previous analysis it follows that in our case
not all modes are reflected, so we have  to explore
whether the superradiant ones propagate to infinity or not.

One can eliminate the first derivative in the radial equation by
introducing the tortoise coordinate
\be
dr_*=\frac{r_0r_+}{\Delta}\, dr,
\end{equation}
leading to
\be
 r_*=\frac1{2\kappa}\ln \left(\frac{r-r_+}{r-r_-}\right),
\end{equation}
where $\kappa=(r_+-r_-)/2r_0r_+$ is the surface gravity of the
horizon. Note that the range of the tortoise coordinate here is
$(-\infty, 0)$ unlike the case of the Kerr metric, where it is the
whole real axis. Rewriting the radial equation as
\be
\frac{d^2R}{dr_*^2} - V R=0,
\end{equation}
we obtain the potential
\be\lb{pot}
 V=\frac{\mu^2 r\Delta}{r_0r_+^2}+ \frac{\Delta l(l+1)}{(r_0r_+)^2}
 -\frac{r^2}{r_+^2}\left(\omega-\frac{am}{r_0r}\right)^2.
\end{equation}
At the horizon $\Delta = 0$ the potential takes the value
\be
 V= - k^2,\quad k=\omega-m\Omega_h,
\end{equation}
where $k$ is the wave frequency with respect to the observer rotating
with the horizon. Therefore radial functions at the horizon are
\be
 R= \e^{\pm ikr_*},
\end{equation}
so the general solution is a superposition of ingoing and outgoing
waves
\be
\psi\sim f(\theta, \varphi)\e^{-i(\omega t+kr_*)}+g (\theta,
\varphi)\e^{-i(\omega t-kr_*)}.
\end{equation}

In the case of the Kerr metric one then constructs the asymptotic
modes as $\exp[i\omega(t\pm r)]$ and finds that there is a constant
flux at infinity of the modes $\omega>0,\, k<0$, i.e. $\omega <
m\Omega_h$ \cite{dW75}. In our case one case see that the
superradiant modes do not propagate to infinity. Indeed, all terms in
the potential (\ref{pot}) containing the black hole parameters
$M,\,a$ are small when $r\to \infty$, so the asymptotic behavior is
still given by Eq. (\ref{asmu}) for $\mu \neq 0$ or (\ref{asmu0}) for
$\mu = 0$. Since all the massive modes, as well as the massless modes
with $\omega<(l+1/2)/r_0$, are confined (do not propagate to
infinity), and since $|m|<l$ and $a < r_+$, leading to $m\Omega_h =
ma/r_0r_+ < (l+1/2)/r_0$, all superradiant modes are reflected.
Therefore there cannot be any superradiant flux at infinity. But it
would be wrong to conclude that there is no superradiance at all. In
fact, these modes are confined in a region outside the black hole. 
Their amplitude will grow up exponentially due to the
amplification of waves impinging on the horizon, leading in the quantum
description to a positive balance between stimulated emission
and absorption. This effect is basically classical in nature 
\cite{Ga68}, so that the rotating dilaton black holes are unstable, 
similarly to the Kerr-AdS spacetime with
reflecting boundary conditions \cite{HaRe99}. 

It is interesting also to analyse the behavior of the Klein-Gordon
modes in the case of a naked singularity, $M < a$. One can show that
there are modes reflected from the singularity, so that the spectrum
is partially discrete. For the discrete spectrum the Klein-Gordon
norm
\be
||\psi||^2 = \int_{\Sigma}j^{\mu}(x)\,d\Sigma_{\mu}
\end{equation}
has to be finite. Choosing the spacelike hypersurface $\Sigma$ to be
a hyperplane $t =$ constant, this condition reads here
\ba
||\psi^2|| & = & \int(\omega
g^{00}-mg^{0\varphi})|\psi|^2\sqrt{|g|}\,d^3x \nonumber\\ & = &
4\pi\int_{0}^{\infty}\left(\omega -\frac{am}{r_0r}\right)\frac{(r_0r)^2}
{\Delta}R^2(r)\,dr <
\infty. \lb{bound1}
\ea
The discussion of the eigenvalue problem depends on the value of the
rotation parameter $a$.

a) For $a = 0$, $M < 0$, the radial wave function behaves near $r =
0$ as
\be
R(r) \sim C_3 + C_4\ln{r}.
\end{equation}
However in this case $\ln{r}$ is a singular solution of the
Klein-Gordon equation (\ref{KG}) (the action of the Klein-Gordon
operator on it yields a delta function), so that the regularity
condition demands $C_4 = 0$. For $\mu = 0$, the solution of the
radial equation which is bounded at infinity is
\be
R(r)=C_1\,(r-2M)^{-1/2-\nu} F\bigg(\frac12+\nu+i\omega r_0
,\frac12+\nu-i\omega r_0,1+2\nu;\frac{-2M}{r-2M}\bigg)\,,
\end{equation}
with $F$ a hypergeometric function. This diverges logarithmically for
$r \to 0$, so that there are no regular normalisable solutions. So in
this case one has again the same continuous spectrum as in the linear
dilaton case, now describing normal scattering by the timelike
singularity $r = 0$. On the other hand,
for $\mu \neq 0$, the normalisability and regularity conditions lead
to the two constraints $C_2 =0$ in (\ref{asmu}) and $ C_4 = 0$, so
that the energy spectrum is discrete. Our numerical computations show
that, at least for small values of the angular momentum $l$, the
low-lying energy levels are approximately evenly spaced.

b) For $a > 0$, $M < a$, the approximate radial equation near $r = 0$
is
\be
a^2R''(r) - \lambda^2R(r) \simeq 0,
\end{equation}
with $\lambda = \sqrt{l(l+1) - m^2}$. This is solved by a
series in powers of $r$. In the present case $R=r$ is a singular
solution of the Klein-Gordon equation (\ref{KG}) (again, the action on it
of the Klein-Gordon operator gives a delta function), so that the
regularity condition now reads
\be\lb{reg}
R'(0) = 0.
\end{equation}
For $\mu \neq 0$ we again have two boundary conditions, $C_2 = 0$ and
(\ref{reg}), leading to a discrete spectrum. For $\mu = 0$, the
energy spectrum contains as before the continuous component $\omega
\in [(l+1/2)/r_0\,,\,+\infty[$, plus possibly a finite number of
discrete eigenvalues in the range $[0\,,\,(l+1/2)/r_0]$.

In this case ($M < a, \mu = 0$), the radial wave function can be
obtained in closed form. After the change of variables
\begin{equation}
r=M+cx\quad\mbox{with}\quad c=\sqrt{a^2-M^2},
\end{equation}
the radial equation (\ref{radial}) becomes
\begin{equation}
\partial_x\left((1+x^2)\partial_x R\right)+\left(-l(l+1)
+\frac{(\omega^{'}+m^{'} x)^2}{1+x^2}\right)R=0 \label{newradial}
\end{equation}
where
\begin{equation}
\omega^{'}=\frac{Mr_0\omega-a m}{c}, \quad m^{'}=r_0\omega.
\end{equation}
Remarkably (\ref{newradial}) is identical to the massless radial
equation in rotating Bertotti-Robinson spacetime (Ref.\cite{conform},
Eq.(3.13) with $\mu=0$), though the boundary conditions are quite
different here. Putting
\be
x \equiv i\frac{\xi+1}{\xi-1},
\end{equation}
(\ref{newradial}) reduces to the hypergeometric type equation
\begin{equation}
\xi^2(\xi-1)^2\partial_\xi^2 R+\xi(\xi-1)^2 \partial_\xi R-l(l+1)\xi
R-\frac{1}{4}(\omega_+^{'}\xi-\omega_-^{'})^2R=0, \label{hypergeom}
\end{equation}
with $\omega_\pm^{'}=\omega^{'}\pm i m^{'}$. The solution of this
equation which is bounded at infinity is
\begin{equation}
R=C_1 \xi^{(\omega'-im')/2}(1-\xi)^{1/2+\nu}
F\bigg(\frac12+\nu+\omega',\frac12+\nu-im',1+2\nu;1-\xi\bigg)
\label{solution}.
\end{equation}
The regularity condition (\ref{reg}) may then be solved numerically
for given values of the quantum numbers ($l,m$). We find that the
number of discrete energy levels is equal to the azimuthal quantum
number $m$, the sign of the energy $\omega$ being opposite to that of $m$.

\section{Stability of the static solution}
\setcounter{equation}{0}

Having established the superradiant classical instability of the
rotating linear dilaton black holes, we now investigate linearization
stability of the static solution (\ref{gstatic})-(\ref{phiFstatic}).
A potentially dangerous perturbation is the $s$-mode. An electric
time-dependent spherically-symmetric solution of the EMDA field
equations may be written as
\ba
ds^2 & = & \e^{2\gamma}dt^2-\e^{2\alpha}dr^2-\e^{2\beta}d\Omega^2, \\
F & = & \frac{r_0}{\sqrt2}\,\e^{\alpha-2\beta+\gamma+2\phi}\,dr\wedge
dt\,,
\ea
where the metric functions $\alpha$, $\beta$, $\gamma$ and the
dilaton $\phi$ depend on $r$ and $t$. We assume that these fields are
small perturbations of the static background fields of
(\ref{gstatic})-(\ref{phiFstatic}),
\ba
& \gamma(r,t)=\gamma_0(r)+\epsilon \gamma_1(r,t),&
\alpha(r,t)=\alpha_0(r)+\epsilon\alpha_1(r,t)\nonumber\\ &
\beta(r,t)=\beta_0(r)+\epsilon\beta_1(r,t),&
\phi(r,t)=\phi_0(r)+\epsilon \phi_1(r,t).
\ea
Choosing the gauge $\beta_1(r,t)=0$ (that is, $\e^{2\beta} = r_0r$), we
obtain for the linearized Klein-Gordon equation and the linearized
Einstein equations for the components $R_2^2$ and $R_{01}$:
\begin{eqnarray}
r_0^2\frac{\ddot{\phi_1}}{(r-2M)^2}-\phi_1^{''}-\frac{2(r-M)}{r(r-2M)}
\phi_1^{'}+\frac{\alpha_1^{'}-\gamma_1^{'}}{2r}
+\frac{\phi_1+\alpha_1}{r(r-2M)} =0, \label{lKG} \\
\frac{\alpha_1^{'}-\gamma_1^{'}}{2r}+\frac{\alpha_1-\phi_1}{r(r-2M)}=0,
\label{R22} \\ \dot{\alpha_1}-\dot{\phi_1}=0 \label{R01},
\end{eqnarray}
where $\dot{}=\partial_t$ and ${}^{'}=\partial_r$. Integrating the
constraint equation (\ref{R01})
\begin{equation}
\alpha_1=\phi_1 \label{alpha1},
\end{equation}
and eliminating the perturbations $\alpha_1$ and $\gamma_1$ between
the equations (\ref{lKG}), (\ref{alpha1}) and (\ref{R22}) we arrive
at the effective Klein-Gordon equation
\begin{equation}
\frac{r_0^2\ddot{\phi_1}}{(r-2M)^2}-\phi_1^{''}-
\frac{2(r-M)}{r(r-2M)}\phi_1^{'}+\frac{2}{r(r-2M)}\phi_1=0.
\end{equation}
The resulting equation for growing modes
$\phi_1(r,t)=\e^{kt}\phi_1(r)$
\begin{equation}
-\phi_1^{''}-\frac{2(r-M)}{r(r-2M)}\phi_1^{'}+
\left(\frac2{r(r-2M)}+\frac{k^2r_0^2}{(r-2M)^2}\right)\phi_1=0,
\end{equation}
is seen to be identical to the radial Klein-Gordon equation
(\ref{radial}) (with $a = 0$) for a massless scalar field with the
particular value of the orbital quantum number $l=1$ and imaginary
frequency $\omega = ik$.

Introducing the new radial variable
\begin{equation}
r_*= \frac{1}{2M}\ln\frac{r-2M}{r},
\end{equation}
we obtain the equation
\begin{equation}\lb{stabeq}
-\phi_{**}+(2r(r-2M) + k^2r_0^2r^2)\phi=0
\end{equation}
For real $k$ the effective potential in (\ref{stabeq}) is positive
definite for all $r > 2M$ (in the black hole case $M > 0$), or for
all $r$ (in the singular case $M \le 0$), so that there are no
bounded solutions. Therefore, the static linear dilaton black holes are
linearly stable in the $s$-sector (and presumably stable with respect
to other modes as well).

\section{Conclusion}
We have constructed new exact solutions to the Einstein-Maxwell-dilaton-axion
system in four dimensions describing black holes which asymptote to
the linear dilaton background. Although these solutions are not asymptotically
flat and break all supersymmetries,
they are interesting since their asymptotic region is  
closely related to an exact solution of string theory.
Non-rotating solutions may be obtained in different ways, the simplest being
to pass to the near-horizon limit of a ``parent'' BPS dilaton black hole.
Using the $Sp(4,R)$ duality of the stationary EMDA system we were also
able to find the rotating (and NUT) generalizations of the above
solutions. The rotating linear dilaton black hole has some similarity 
with the Kerr one, but is very different asymptotically. In the
limit of a vanishing mass parameter the rotating holes tend to the 
supersymmetric dilaton-axion generalization of the
Israel-Wilson-Perj\`es solutions.

Rather unexpectedly, our new black holes have a very simple counterpart in 
six dimensions,
where they can be uplifted using a non-local procedure involving dualization.
Actually it is the smeared vacuum five-dimensional black hole with two equal
rotation parameters. The spacetime symmetry of the six-dimensional solutions 
has a component inherited from the internal symmetry of the stationary 
EMDA system.

The physical mass and angular momentum were computed using the Brown-York 
definition
of quasilocal gravitational energy-momentum tensor, and shown to satisfy the
first law of thermodynamics consistent with the definition of the Hawking
temperature via the surface gravity and of the geometric entropy as the quarter
of the horizon area. A remarkable feature of the static linear dilaton black
hole is that the
temperature is constant (depending on a parameter of the background) and the
mass is independent of it, so the heat capacity is zero. The rotating black hole possesses an ergosphere with
an associated superradiance phenomenon. Analysing the asymptotic behavior
of the Klein-Gordon modes, one finds that the superradiant modes
do not propagate to infinity, but are confined in some region outside the
black hole. This leads to an exponential growth of the superradiant modes,
equivalent to
classical instability of the rotating solution. The remaining  static
black hole is shown to be stable with respect to spherical
perturbations.
We have also found that in the case of
a naked singularity the Klein-Gordon operator has regions of
discrete spectrum. 
It is expected that ten-dimensional brane generalizations of our solutions
exist as well, and presumably have holographic interpretation as 
thermal states of the little string theory.

\section*{Acknowledgements}
One of the authors (DG) would like to thank J. Lemos for
useful discussions. He also thanks LAPTH (Annecy) and CENTRA IST (Lisbon)
where parts of the paper have been written for hospitality and support.
The support by the RFBR under grant 00-02-16306 is acknowledged.



\newpage

\begin{figure}
\centerline{\epsfxsize=200pt\epsfbox{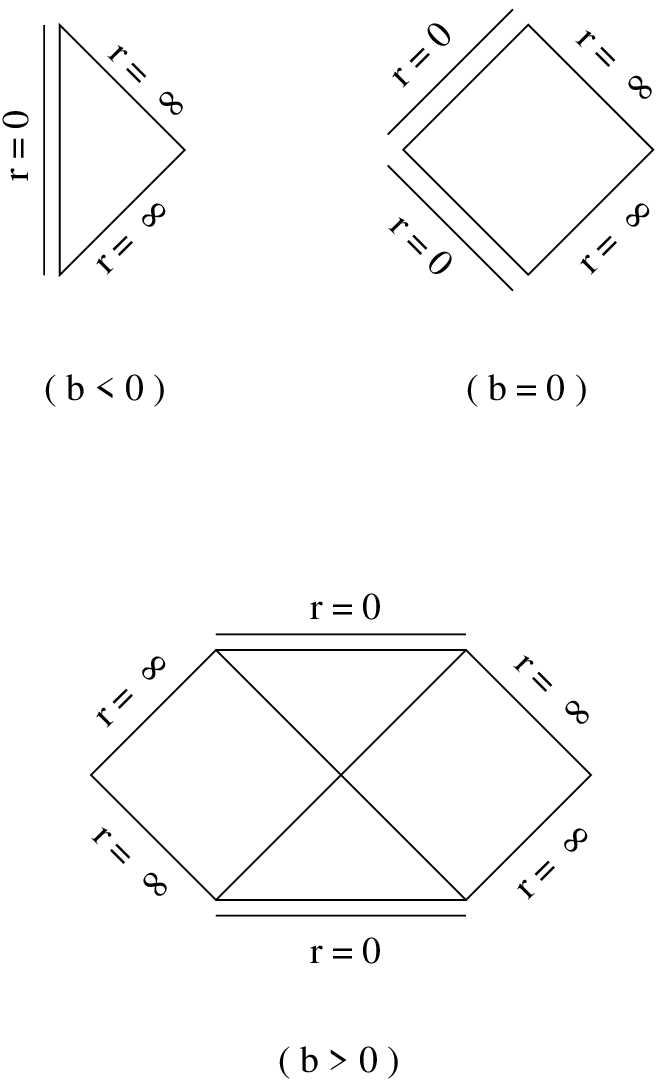}}
\vspace{2cm}
\caption{\vspace{5cm} Penrose diagrams of the static spacetimes (\ref{gstatic}) 
with $b<0$, $b=0$ and $b>0$.}
\end{figure}

\end{document}